\documentclass[%
reprint,
 amsmath,amssymb,
 aps,
]{revtex4-2}

\usepackage{amsmath,amssymb, bm}
\usepackage{graphicx}% Include figure files
\usepackage{dcolumn}% Align table columns on decimal point
\usepackage{bm}% bold math
\usepackage{physics}
\usepackage{comment}
\usepackage{ulem}
\usepackage{xcolor}
\usepackage{hyperref}% add hypertext capabilities
\usepackage{xifthen}% provides \isempty test
\newcommand{\vq}[2][]{                % vq = Vector of Quadratures
  \ifthenelse{\isempty{#1}}           %
    { \hat{\pmb{#2}} }                % if #2 is empty
    { \hat{\pmb{#2}}_\mathrm{#1} }    % if #2 is not empty
}
\newcommand{\vqtil}[2][]{             % vq = Vector of Quadratures, til = Tilde
  \ifthenelse{\isempty{#1}}           %
    { \hat{\til{\pmb{#2}}} }                % if #2 is empty
    { \hat{\til{\pmb{#2}}}_\mathrm{#1} }    % if #2 is not empty
}
\newcommand{\tq}[2][]{                % tq = matrix of Tranformation of Quadratures
  \ifthenelse{\isempty{#1}}           %
    { \mathbb{#2} }                   % if #2 is empty
    { \mathbb{#2}_\mathrm{#1} }       % if #2 is not empty
}
\newcommand{\vs}[2][]{                % vq = Vector of Scalars
  \ifthenelse{\isempty{#1}}           %
    { \mathbf{#2} }                   % if #2 is empty
    { \mathbf{#2}_\mathrm{#1} }       % if #2 is not empty
}

\newcommand{\Tq}[3][]{                % tq = matrix of Tranformation of Quadratures
  \ifthenelse{\isempty{#1}}           %
    { \mathbb{#3} }                   % if #2 is empty
    { \mathbb{#3}^{\rm #1}_{\rm #2} }       % if #2 is not empty
}

\newcommand{\mr}[1]{\mathrm{#1}}

\date{\today}

\begin{document}
\preprint{APS/123-QED}
\title{Teleportation-based Speed Meter for Precision Measurement}% Force line breaks with 
\author{Yohei Nishino$^{1,2}$}
 \email{yohei.nishino@grad.nao.ac.jp}
\author{James W. Gardner$^{3,4}$}
\author{Yanbei Chen$^{4}$}
\author{Kentaro Somiya$^{5}$}

\affiliation{$^1$Department of Astronomy, University of Tokyo, Bunkyo, Tokyo 113-0033, Japan,}
\affiliation{$^2$Gravitational Wave Science Project, National Astronomical Observatory of Japan (NAOJ), Mitaka City, Tokyo
181-8588, Japan,}
\affiliation{$^3$OzGrav-ANU, Centre for Gravitational Astrophysics, Research Schools of Physics, and of Astronomy and Astrophysics, The Australian National University, Canberra, ACT 2601, Australia}
\affiliation{$^4$Walter Burke Institute for Theoretical Physics, California Institute of Technology, Pasadena, California 91125, USA} 
\affiliation{$^5$Department of Physics, Institute of Science Tokyo, Meguro, Tokyo, 152-8551, Japan}

\date{\today}% It is always \today, today,

\begin{abstract}
We propose a quantum teleportation-based speed meter for interferometric displacement sensing. Two equivalent implementations are presented: an online approach that uses real-time displacement operation and an offline approach that relies on post-processing. Both implementations reduce quantum radiation pressure noise and surpass the standard quantum limit of measuring displacement. We discuss potential applications to gravitational-wave detectors, where our scheme enhances low-frequency sensitivity without requiring modifications to the core optics of a conventional Michelson interferometer (e.g., substrate or coating properties). This approach offers a new path to back-action evasion enabled by quantum entanglement.
\end{abstract}

\maketitle
\section*{Introduction}

Precision measurements in cavity optomechanics are fundamentally limited by quantum uncertainties. In free-mass displacement measurements, Heisenberg’s uncertainty principle imposes the standard quantum limit (SQL)~\cite{1968JETP.26.831}, arising from a trade-off between measurement precision (associated with position uncertainty) and back-action (associated with momentum uncertainty). Improving measurement precision inevitably leads to an increase in back-action noise.

Various back-action evasion techniques have been developed to mitigate this noise, such as squeezing injection~\cite{Meystre1983QuantumOE, PhysRevD.23.1693, PhysRevD.30.2548, Jaekel_1990, PhysRevD.65.022002}, variational readout~\cite{PhysRevD.65.022002, VYATCHANIN1995269}, effective negative-mass systems~\cite{Julsgaard_2001, PhysRevLett.102.020501, PhysRevA.87.063846, PhysRevLett.105.123601, PhysRevLett.117.140401, M_ller_2017, PhysRevLett.121.031101}, and optical-spring techniques~\cite{PhysRevD.65.042001}.  Among these, squeezed-state injection has become well-established in gravitational-wave detectors (GWDs). By squeezing a specific quadrature of the vacuum field, one can reduce either position or momentum uncertainty at the expense of the other. The Laser Interferometer Gravitational-wave Observatory (LIGO) has demonstrated an improvement of approximately 3~dB at around 50 Hz, exceeding the SQL~\cite{Yu_2020,Jia_2024}.

An alternative way to beat the SQL---without relying on squeezing---is to perform quantum non-demolition (QND) measurements of observables that commute with themselves at different times~\cite{doi:10.1126/science.209.4456.547,1995qume.book.....B}. For a free mass, the momentum operator is one such operator and thus can be monitored in a QND manner that inherently beats the SQL without squeezing. Since momentum is proportional to velocity, such a device that monitors the momentum is often called a speed meter~\cite{PhysRevD.61.044002, BRAGINSKY1990251}. 

This work proposes a protocol to realize a speed meter through quantum teleportation. The principle of speed measurement is depicted in Fig.~\ref{fig:diagram}a; the probe laser is coupled the the object mass twice canceling out the back action of the measurement (see section 4.2. in Ref~\cite{Danilishin_2012}). For interferometric sensors, previous work realized this coupling using the Sagnac interferometer~\cite{Chen_2003}, sloshing cavity mode~\cite{PhysRevD.66.122004} or manipulation of polarization~\cite{2018LSA.....7...11D,KNYAZEV20182219}.

Quantum teleportation is widely regarded as a foundational technology for future quantum networks~\cite{PhysRevLett.81.5932,Duan_2001} and distributed quantum computing systems~\cite{1999Natur.402..390G}. The reliable transfer of unknown quantum states between distant locations enables large-scale quantum communication protocols and significantly enhances data security~\cite{Bennett_2014,PhysRevLett.67.661}. Quantum teleportation has also been proposed to extend the baseline of astrophysical telescopes~\cite{PhysRevLett.109.070503}.

In this paper, we propose applying quantum teleportation to quantum back-action evasion by means of speed meter.
A motivation of our scheme is similar to the conditional or teleportation broadband squeezing approaches proposed in~\cite{Ma_2017,PhysRevA.110.022601}, which aim to circumvent technical difficulties in GWDs by exploiting quantum entanglement. Those proposals eliminate the need for large-scale, low-loss, high-finesse filter cavities for broadband squeezing. Our proposal uses quantum entanglement to convert interferometric displacement sensors into speed meters without modifying the substrate or coating properties of the interferometer's core optics. This approach thus benefits both current~\cite{Somiya_2012,Acernese_2015,Aasi_2015} and next-generation gravitational-wave detectors~\cite{Hild_2011,Abbott_2017}.

%The rest of this paper is structured as follows. In Sec.~\ref{sec.2}, we review the fundamental principle of the double-pass speed meter. In Sec.~\ref{sec.3}, we describe our proposed scheme for implementing a quantum teleportation-based speed meter. Then, in Sec.~\ref{sec.4}, we present a sensitivity analysis for displacement measurement employing our scheme and consider the gravitational-wave detector application. Finally, in Sec.~\ref{sec.5}, we discuss the implications of our findings, and Sec.~\ref{sec.6} concludes the paper.

\section*{Results}
\subsection*{Principle of Speed Measurement}\label{sec.2}
We briefly review the principles of the speed meter. In conventional position measurements, the detector's sensitivity is ultimately limited by quantum noise, which comprises quantum radiation pressure noise (QRPN) and shot noise. The standard quantum limit (SQL) is given by
\begin{align}
    x_\mr{SQL} (\Omega) = \sqrt{\frac{2\hbar}{m\Omega^2}},
\end{align}
where \(\Omega\) is the angular frequency, \(\hbar\) is Planck's constant, and \(m\) is the mass of the test object. The SQL originates from the non-commutativity of the position operators at different times; that is, the position operator of the test mass at time \(t\) does not commute with the one at a later time \(t^\prime\):
\begin{align}
    [\hat{x}(t),\ \hat{x}(t^\prime)] \neq 0.
\end{align}
Consequently, there is an inherent trade-off between QRPN and shot noise, which prevents both noise sources from being simultaneously reduced.

In contrast, observables that commute at different times can be monitored with arbitrary precision. These are referred to as QND observables. The momentum of a free mirror is one example of a QND observable. For a freely evolving mass, its momentum is proportional to its speed, i.e.\ $\hat{p} = m\hat{v}$, which is why a momentum meter is also called a ``speed meter." However, when the mirror is coupled to a probe such as light, a more careful analysis reveals that the canonical momentum is no longer simply proportional to the speed. Instead, it takes the form
\begin{align}
    \hat{p} = m\hat{v} - \alpha\hat{a}_1,
\end{align}
where \(\hat{a}_1\) is the amplitude quadrature of the probe light and \(\alpha\) represents the optomechanical coupling strength. Nevertheless, it is still possible to reduce the back-action noise below the SQL by exploiting the correlation between the phase and amplitude quadratures of the outgoing field (see Section 2.11 of \cite{HaixingPhD2010} and Section 4.5.2 of \cite{Danilishin_2012}).

\begin{figure}
    \centering
    \includegraphics[width=0.9\linewidth]{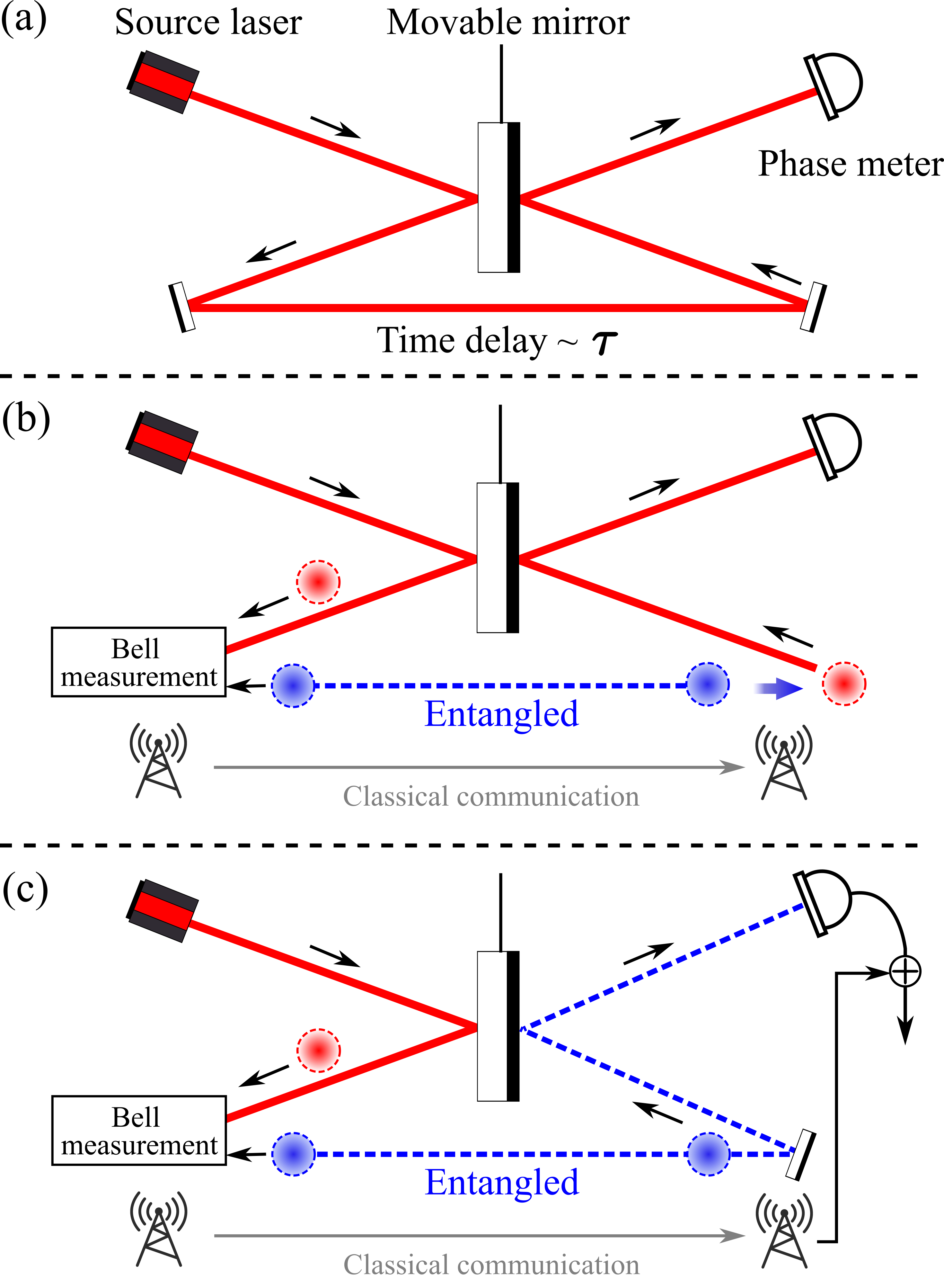}
    \caption{(a) Simplified model of the speed meter as presented in Ref.~\cite{Danilishin_2012}. (c) Model illustrating the online approach, and (d) model for the offline approach of the teleportation-based scheme.}
    \label{fig:diagram}
\end{figure}

An illustration of the light--mirror interaction is provided in Fig.~\ref{fig:diagram}a (also see Ref.~\cite{Danilishin_2012}). Light from the source laser impinges on the front side of the movable mirror, is reflected, and then recycled to the back side of the mirror with a time delay \(\tau\). At the second interaction (at time \(t+\tau\)), the intensity fluctuations of the light push the mirror in the opposite direction compared to the first interaction (at time \(t\)), thereby canceling the initial radiation pressure force. Consequently, the observable measured by the phase meter becomes
\begin{align}
\phi \propto \hat{x}(t+\tau) - \hat{x}(t) \sim \tau\,\bar{\hat{v}},
\end{align}
where \(\bar{v}\) denotes the average velocity of the mirror. Note that when the light interacts with the mirror inside optical cavities, the delay \(\tau\) corresponds to the cavity storage time.

\subsection*{Nonreciprocal Coupling via Teleportation}\label{sec.3}
\begin{figure}
    \centering
    \includegraphics[width=0.85\linewidth]{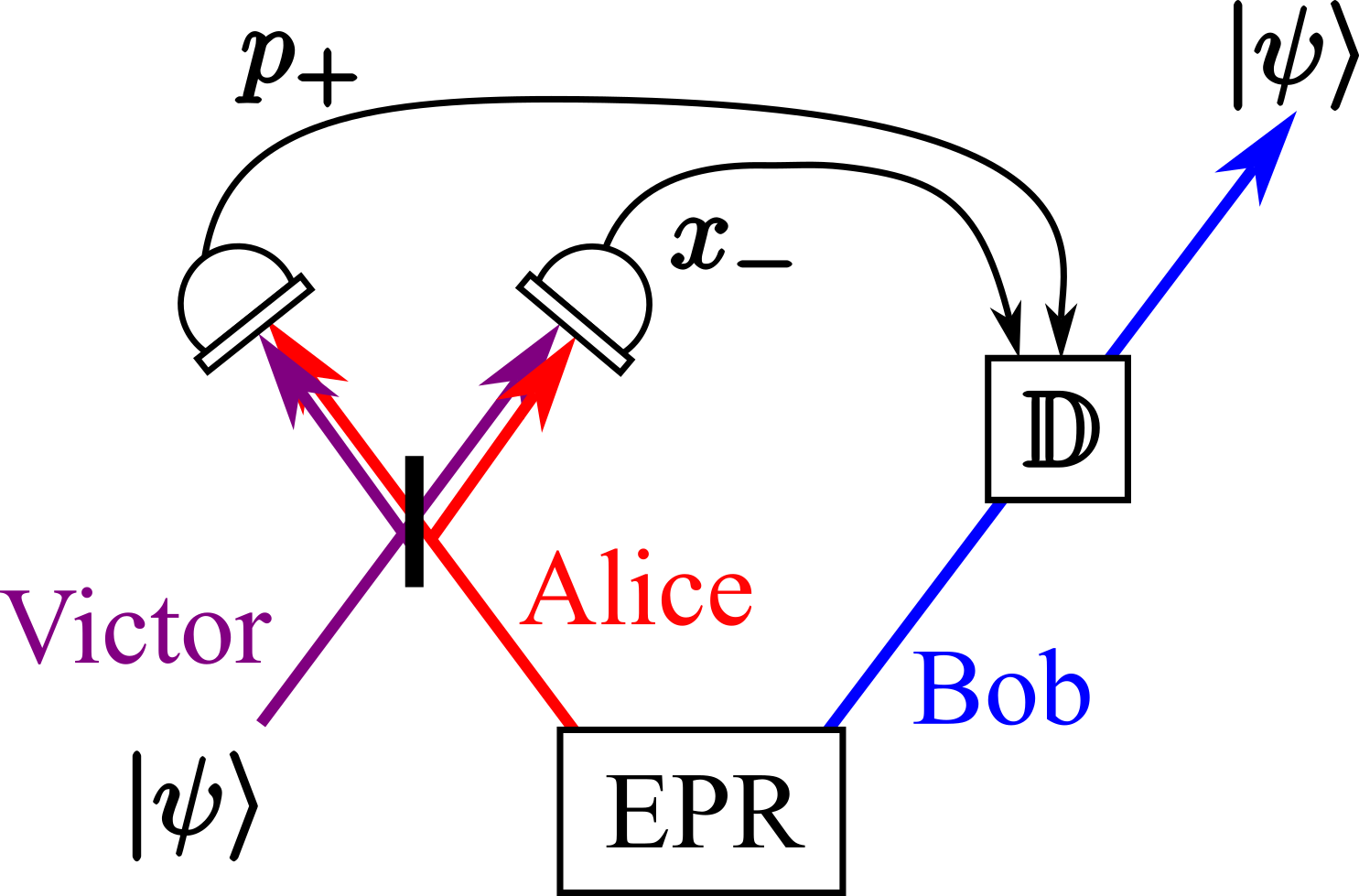}
    \caption{Diagram of continuous-variable teleportation. Alice and Bob share an EPR-entangled state, and Victor prepares the target state $\ket{\psi}$. A Bell measurement on Victor's state and one of Alice’s modes yields outcomes $x_-$ and $p_+$, which Bob uses to displace his mode. As a result, Victor’s state is teleported to Bob.}
    \label{fig:teleportation}
\end{figure}

In this section, we discuss how to achieve a speed meter using teleportation. Our teleportation scheme is based on the Braunstein-Kimble protocol~\cite{PhysRevLett.80.869}. In this protocol, the target state (hereafter referred to as \textit{Victor}) is teleported from one half of an EPR pair (referred to as \textit{Alice}) to the other half (referred to as \textit{Bob}) via the transmission of classical information (see Fig.~\ref{fig:teleportation}). The transmitted information is obtained from a joint (Bell) measurement performed on Alice and Victor. In an ideal EPR state, two modes are completely correlated, while the individual modes remain completely uncertain (i.e., they exhibit infinite EPR noise). In a conventional homodyne measurement, measuring one observable typically demolishes information about its conjugate; however, when the highly noisy EPR state is mixed with Victor's state prior to measurement, the measurement reveals no information about Victor's state. Due to the perfect quantum correlations in an ideal EPR state, only Bob can reconstruct Victor's state using the classical information from Alice.

%In the following, we first describe the method for quantum state preparation and outline the system dynamics using Hamiltonian modeling in a general form. Then, we propose two approaches for speed measurement—namely, the online and offline approaches—and demonstrate their equivalence.

In Fig.~\ref{fig:diagram}b, an Einstein-Podolsky-Rosen (EPR) entangled pair of photons is distributed to stations on the front and back sides of a mirror, which we denote as Alice and Bob, respectively. The photon that interacts with the mirror on the front side is labeled Victor. Instead of recycling Victor to the mirror's back side, Alice performs a joint (Bell) measurement on her photon and Victor's photon. The output of this Bell measurement is a combination of the two quadratures of the input fields necessary for teleportation.

Alice then transmits the measurement results to Bob through classical communication. Bob applies a displacement operation to his photon according to the received data, thereby teleportating Victor's state to Bob. The teleported state is subsequently directed to the mirror from its back side, where it interacts with the mirror once more. With perfect fidelity of teleportation, this process is therefore equivalent to that depicted in Fig.~\ref{fig:diagram}a.
%(can be achieved only when $r\rightarrow\infty$)

In the following, we first outline the preparation of the initial states for teleportation, then present two equivalent approaches—the online and offline methods, shown in panels (b) and (c) of Fig.~\ref{fig:diagram}, respectively. Then, the system dynamics are analyzed using a general Hamiltonian model. %The scheme is applied to an interferometric sensor in a later section.

\subsubsection*{State Preparation and System Dynamics}
%In this section, we begin by outlining a method to create non-reciprocal interactions via quantum teleportation. Two equivalent approaches are proposed, the online and offline approach, which are shown in panels b and c of Fig.~\ref{fig:diagram}, respectively. First, we consider the online model, which provides a qualitative description. We then demonstrate that the offline approach is equivalent to the online model.

\begin{figure}
    \centering
    \includegraphics[width=0.8\linewidth]{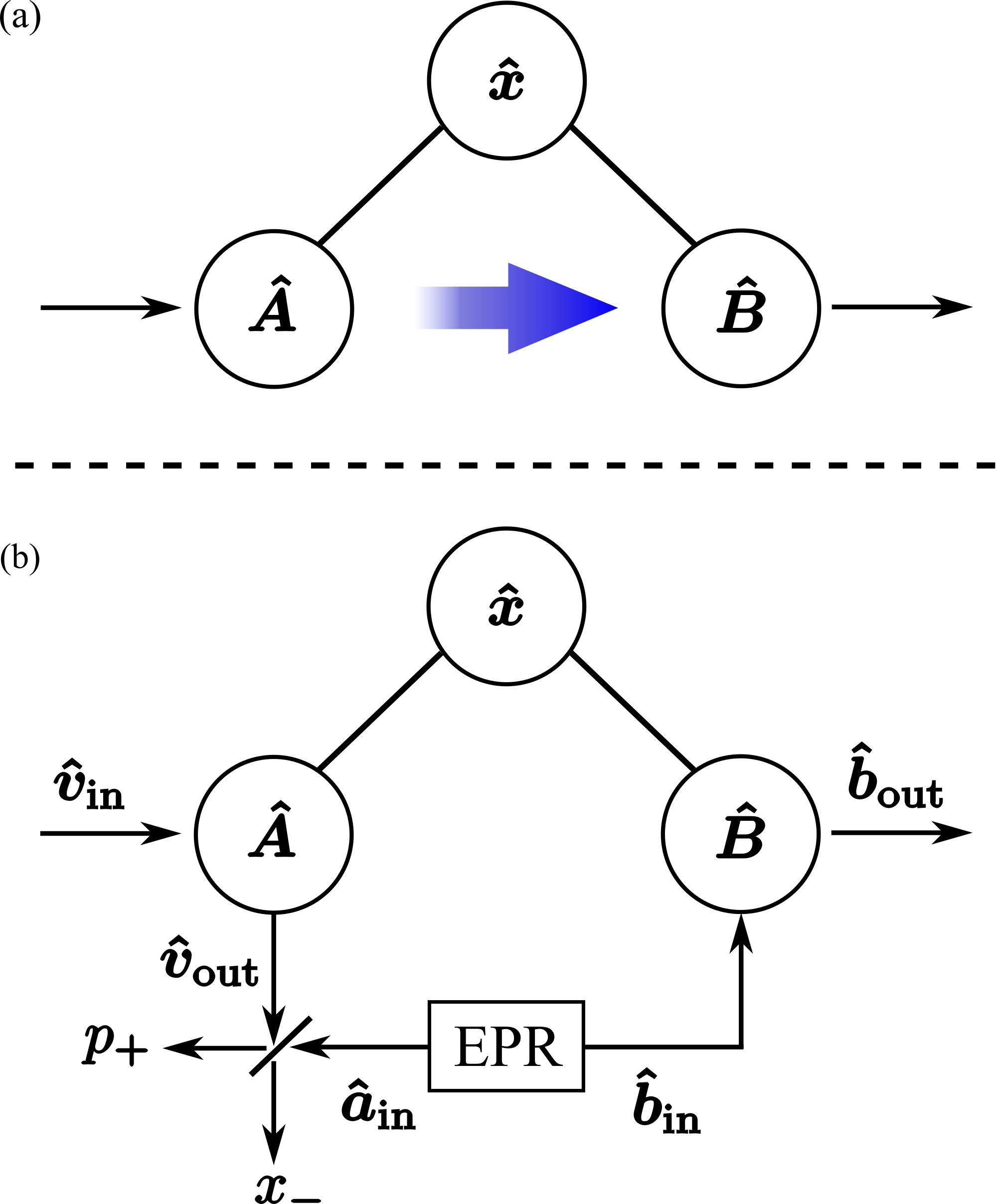}
    \caption{(a) A mode diagram of the speed meter. (b) Input and output relation.}
    \label{fig:mode_diagram}
\end{figure}

Throughout this work, we adopt a quadrature representation based on the two-photon formalism~\cite{PhysRevA.31.3093,PhysRevA.31.3068}. Given that \(\hat{a}_\omega\) denotes the single-photon annihilation operator of the mode with frequency \(\omega\), the two-photon quadrature amplitudes are defined as
\begin{align}
    \hat{a}_{1,\Omega} &= \frac{\hat{a}_{\omega_0+\Omega}+\hat{a}^\dagger_{\omega_0-\Omega}}{\sqrt{2}}, \\
    \hat{a}_{2,\Omega} &= \frac{\hat{a}_{\omega_0+\Omega}-\hat{a}^\dagger_{\omega_0-\Omega}}{i\sqrt{2}},
\end{align}
where \(\omega_0\) is the carrier frequency and \(\Omega\) is the sideband frequency. We denote the fields associated with Victor, Alice, and Bob as
\begin{align}
    \pmb{\hat{a}}=\begin{pmatrix}\hat{a}_{1,\Omega}\\ \hat{a}_{2,\Omega}\end{pmatrix},\quad 
    \pmb{\hat{b}}=\begin{pmatrix}\hat{b}_{1,\Omega}\\ \hat{b}_{2,\Omega}\end{pmatrix},\quad 
    \pmb{\hat{v}}=\begin{pmatrix}\hat{v}_{1,\Omega}\\ \hat{v}_{2,\Omega}\end{pmatrix}.
\end{align}

For brevity, we omit the explicit \(\Omega\) dependence hereafter and introduce subscripts ``in'' and ``out'' to denote input and output fields, for example, \(\pmb{\hat{a}}_\mr{out} = \{\hat{a}_{1,\mr{out}},\,\hat{a}_{2,\mr{out}}\}^T\). 

The EPR entanglement, which is a two-mode squeezed vacuum state, is characterized in terms of the quantum spectral density of the four EPR operators \(\bigl(\hat{a}_{j,\mr{in}}\pm\hat{b}_{j,\mr{in}}\bigr)/\sqrt{2}\) for $j=1,2$ as~\cite{2000_Duan}
\begin{align}
    S_{(\hat{a}_{j,\mr{in}}\pm\hat{b}_{j,\mr{in}})/\sqrt{2}} &= e^{\mp (-1)^j 2r},\quad 
    % S_{(\hat{a}_{1,\mr{in}}\pm\hat{b}_{1,\mr{in}})/\sqrt{2}} &= e^{\pm 2r},\quad 
    % S_{(\hat{a}_{2,\mr{in}}\pm\hat{b}_{2,\mr{in}})/\sqrt{2}} = e^{\mp 2r},
    \label{Eq:Spectrum}
\end{align}
where \(r\) is the squeezing factor. When $r\rightarrow\infty$, the noise spectra of $S_{(\hat{a}_{1}-\hat{b}_{1})/\sqrt{2}}$ and $S_{(\hat{a}_{2}+\hat{b}_{2})/\sqrt{2}}$ approach zero, corresponding to the original EPR entanglement~\cite{PhysRev.47.777}.

%The interferometer is pumped by two lasers operating at the same frequencies as Victor’s and Bob’s fields (details provided later). 
The input field for Victor, \(\pmb{\hat{v}}_\mr{in}\) (a pure vacuum state, depicted in purple), is injected to mode $\hat{A}$, interacts with the pump, and then exits as \(\pmb{\hat{v}}_\mr{out}\). This field enters the Bell measurement and results in measurement outcomes as follows
\begin{align}
    \pmb{\alpha} = \frac{1}{\sqrt{2}}
    \begin{pmatrix}
       v_{1,\mr{out}}-a_{1,\mr{in}}\\[1mm]
       v_{2,\mr{out}}+a_{2,\mr{in}}
    \end{pmatrix}
    \equiv
    \begin{pmatrix}
       x_{-}\\[1mm]
       p_{+}
    \end{pmatrix}. \label{Eq:alpha}
\end{align}

We now describe the system dynamics using an analytic Hamiltonian formalism.
Fig.~\ref{fig:mode_diagram}a illustrates the mode diagram of the non-reciprocal speed measurement. Two cavity modes, \(\hat{A}\) and \(\hat{B}\), are coupled to the position of the oscillator's mechanical mode \(\hat{x}\). The interaction between \(\hat{A}\) and \(\hat{B}\) is non-reciprocal: the outgoing field from \(\hat{A}\) enters \(\hat{B}\), while the reverse process does not take place. As a result, the input field entering \(\hat{A}\) exits from \(\hat{B}\) after interacting with the mirror twice. By inverting the sign of the two terms in the Hamiltonian related to the optomechanical interaction, the measurement's back-action can be effectively canceled out.

The input and output relation of the fields and modes is depicted in Fig.~\ref{fig:mode_diagram}b.
The Hamiltonian of the system is given by~\cite{PhysRevA.51.2537,Chen_2013}
% \begin{widetext}
\begin{align}
\hat{H} &=  \hbar\,\omega_\mr{a} \Bigl(\hat{A}+\hat{A}^\dagger\Bigr)
\Bigl(\hat{x}-x\Bigr) \label{eq:Hamiltonian}
\\&-\hbar\,\omega_\mr{a} \Bigl(\hat{B}+\hat{B}^\dagger\Bigr)\nonumber
\Bigl(\hat{x}-x\Bigr)
\\&+\frac{1}{2}\left(\frac{\hat{p}^2}{m}+m\,\omega_\mr{m}^2\,\hat{x}^2\right),\nonumber
\end{align}
% \end{widetext}
where \(x\) is the displacement by external classical force, \(m\) is the mirror mass, and \(\omega_\mr{m}\) and \(\hat{p}\) denote the natural frequency and momentum operator of the mechanical mode, respectively. \(\omega_\mr{a}\) is the optomechanical coupling constant between the mechanical and cavity modes.
In deriving Eq.~\eqref{eq:Hamiltonian}, we have applied the rotating-wave approximation to ignore the self-evolution of the light modes.

The Heisenberg-Langevin equations of motion in the time domain are thus:
\begin{align}
\frac{\mr{d}\hat{A}}{\mr{d}t} &= -\gamma\, \hat{A} 
  + \sqrt{2\gamma}\,\hat{v}_\mr{in} - i\omega_\mr{a}\Bigl(\hat{x}-x\Bigr), \notag\\[1mm]
\frac{\mr{d}\hat{B}}{\mr{d}t} &= -\gamma\, \hat{B} 
  + \sqrt{2\gamma}\,\hat{b}_\mr{in} + i\omega_\mr{a}\Bigl(\hat{x}-x\Bigr), \notag\\[1mm]
\frac{\mr{d}\hat{x}}{\mr{d}t} &= \frac{\hat{p}}{m}, \notag\\[1mm]
\frac{\mr{d}\hat{p}}{\mr{d}t} &= -\hbar\,\omega_\mr{a}\Bigl(\hat{A}+\hat{A}^\dagger-\hat{B}-\hat{B}^\dagger\Bigr)
  - m\,\omega_\mr{m}^2\,\hat{x}\,. \label{Eq:EoM}
\end{align}
In the Fourier domain and expressed in terms of quadrature operators, these equations become:
\begin{align}
    -i\Omega \hat{A}_1 &= -\gamma\, \hat{A}_1 + \sqrt{2\gamma}\,\hat{v}_{1,\mr{in}}, \notag\\[1mm]
    -i\Omega \hat{A}_2 &= -\gamma\, \hat{A}_2 + \sqrt{2\gamma}\,\hat{v}_{2,\mr{in}}
    -\sqrt{2}\,\omega_\mr{a}\Bigl(\hat{x}-x\Bigr), \notag\\[1mm]
    -i\Omega \hat{B}_1 &= -\gamma\, \hat{B}_1 + \sqrt{2\gamma}\,\hat{b}_{1,\mr{in}}, \notag\\[1mm]
    -i\Omega \hat{B}_2 &= -\gamma\, \hat{B}_2 + \sqrt{2\gamma}\,\hat{b}_{2,\mr{in}}
    -\sqrt{2}\,\omega_\mr{a}\Bigl(\hat{x}-x\Bigr), \notag\\[1mm]
    -i\Omega \hat{x} &= \frac{\hat{p}}{m}, \notag\\[1mm]
    -i\Omega \hat{p} &= -\sqrt{2}\,\hbar\,\omega_\mr{a}\Bigl(\hat{A}_1-\hat{B}_1\Bigr)
    - m\,\omega_\mr{m}^2\,\hat{x}, \label{Eq:EoM2}
\end{align}
where $\gamma$ is the cavity's bandwidth.
Here,  for \(X=(\hat{A},\hat{B})\), we define the quadratures as
\begin{align}
    X_1 = \frac{X+X^\dagger}{\sqrt{2}},\quad
    X_2 = \frac{X-X^\dagger}{i\sqrt{2}},
\end{align}
and denote \(\pmb{X}=\{X_1, X_2\}^T\).

The output fields of the two cavity modes are given by
\begin{align}
    \pmb{\hat{v}}_\mr{out} &= -\pmb{\hat{v}}_\mr{in}+\sqrt{2\gamma}\,\pmb{\hat{A}}, \notag\\
    \pmb{\hat{b}}_\mr{out} &= -\pmb{\hat{b}}_\mr{in}+\sqrt{2\gamma}\,\pmb{\hat{B}},
\end{align}
which are the standard input-output relations.

\subsubsection*{Online Approach}
In the ``online'' approach, the displacement operation is applied to the light field in real time. Here, we first analyze the direct implementation as it provides clearer physical intuition and directly corresponds to the schematic in Fig.~\ref{fig:diagram}b. We will later show that this operation can be replaced with post-processing (which we call the ``offline'' approach).

Bob's field, denoted as $\pmb{\hat{v}}_\mr{out}$, is displaced based on the Bell measurement's outcomes \(\{x_{-},\,p_{+}\}^T\) and then injected into the mode $\hat{B}$. The field subsequently couples with $\hat{x}$, which cancels the back-action, and then emerges as the output as $\pmb{\hat{b}}_\mr{out}$. %(Note that the beam splitters combining the three beams are frequency-dependent; for example, a triangular optical cavity can be used to realize this.)

The displacement operation, based on the Bell measurement Eq.~\eqref{Eq:alpha}, is expressed as
\begin{align}
     \tq{D}\bigl[\pmb{\hat{b}}_\mr{in};\, x_-, p_+\bigr]
    = \pmb{\hat{b}}_\mr{in}+\sqrt{2}\,\pmb{\alpha}=
    \begin{pmatrix}
       \hat{b}_{1,\mr{in}} + x_-\\[1mm]
       \hat{b}_{2,\mr{in}} + p_+
    \end{pmatrix}. \label{Eq:displace}
\end{align}
Using Eq.~\eqref{Eq:Spectrum}, Eq.~\eqref{Eq:displace} can be rewritten as
\begin{align}
    \pmb{\hat{b}}_\mr{in} = \sqrt{2}\,e^{-r}\,\pmb{\hat{z}} + \pmb{\hat{v}}_\mr{out},
\end{align}
where the operators \(\hat{z}_{1,2}\) have unit variance (i.e., \(S_{\hat{z}_1}=S_{\hat{z}_2}=1\)) and we define \(\pmb{\hat{z}}=\{\hat{z}_1,\hat{z}_2\}^T\).

The output field from mode \(\hat{B}\) is then given by
\begin{align}
    \pmb{\hat{b}}_\mr{out} = \tq[v]{T}\,\pmb{\hat{v}}_\mr{in}+\sqrt{2}\,e^{-r}\,\tq[z]{T}\,\pmb{\hat{z}}+\frac{x}{x_\mr{SQL}}\,\pmb{t}, \label{Eq:b_out}
\end{align}
where
\begin{align}
    \tq[v]{T} &= e^{4i\beta}\begin{pmatrix}
        1 & 0 \\
        -\mathcal{K}_\mr{sm} & 1
    \end{pmatrix},\\[1mm]
    \tq[z]{T} &= e^{2i\beta}\begin{pmatrix}
        1 & 0 \\
        -\mathcal{K}_z & 1
    \end{pmatrix}, \\[1mm]
    \pmb{t} &= e^{2i\beta}\begin{pmatrix}
        0 \\
        \sqrt{2\mathcal{K}_\mr{sm}}
    \end{pmatrix}.
\end{align}
Here, \(\beta = \arctan\frac{\Omega}{\gamma}\) is a phase rotation common to both quadratures, and \(\mathcal{K}_\mr{sm}\) and \(\mathcal{K}_z\) are the optomechanical coupling factors for the speed meter and the auxiliary field \(\pmb{\hat{z}}\), respectively:
\begin{align}
    \mathcal{K}_\mr{sm} &= \frac{16\hbar\,\omega_\mr{a}^2\,\gamma}{m(\gamma^2+\Omega^2)^2},\\[1mm]
    \mathcal{K}_z &= \frac{\gamma+i\Omega}{2i\Omega}\,\mathcal{K}_\mr{sm}.
\end{align}

The readout is performed by projecting the outgoing field onto the homodyne vector \(\pmb{H}^T_{\phi}=\{\sin\phi,\,\cos\phi\}\):
\begin{align}
    \hat{b}_\phi = \pmb{H}^T_{\phi}\,\pmb{\hat{b}}.
\end{align}
The spectral density for displacement measurement (displacement sensitivity) is then
\begin{align}
    S_x(\Omega) = x^2_\mathrm{SQL}\,\frac{\sum_{\mu} \pmb{H}_{\phi}^T\,\tq[\mu]{D}\,\tq[\mu]{S}^\mathrm{in}\,\tq[\mu]{D}^\dagger\,\pmb{H}_{\phi}}{\bigl|\pmb{H}_{\phi}^T\,\pmb{t}\bigr|^2}, \label{eq:Sx}
\end{align}
where $\mu\in\{v,z\}$. The quantum spectral density of the input fields, $\tq[\mu]{S}^\mathrm{in}$, is defined by
\begin{align}
    2\pi\,&\delta(\Omega-\Omega')\,\tq[\mu,\mathit{ij}]{S}^\mathrm{in}(\Omega) \notag\\
    &\equiv \tfrac{1}{2}\langle \mathrm{in}\mid \hat{b}_{i,\mr{in}}(\Omega)\,\hat{b}_{j,\mr{in}}^\dagger(\Omega') + \hat{b}_{j,\mr{in}}^\dagger(\Omega')\,\hat{b}_{i,\mr{in}}(\Omega)\mid \mathrm{in}\rangle,
\end{align}
with $(i,j)=\{1,2\}$ (see Refs.~\cite{Danilishin_2012,2018LSA.....7...11D} for more details).
For a readout angle of \(\phi=\pi/2\) and in the limit \(r\rightarrow \infty\), the displacement sensitivity of the quantum noise–limited interferometer is given by
\begin{align}
    S_{x,\pi/2}^\mr{sm} = \frac{x_\mr{SQL}^2}{2}\Bigl(\frac{1}{\mathcal{K}_\mr{sm}}+\mathcal{K}_\mr{sm}\Bigr). \label{Eq:Spec}
\end{align}
%Conversion between position $x$ and gravitational--wave strain $h$ can be obtained with the length of the cavity $L$ as:
%\begin{align}
%    x = \frac{hL}{2},
%\end{align}
%thus the strain sensitivity is calculated as:
%\begin{align}
%    S_h(\Omega) = \frac{4 S_x(\Omega)}{L^2}.
%\end{align}
Eq.~\eqref{Eq:Spec} agrees with the sensitivity shown in Ref.~\cite{2018LSA.....7...11D}.

\subsubsection*{Offline Approach}
The online displacement operation can be replaced by post-processing, which is experimentally easier as it can be implemented ``offline".
In this case, the input–output relation can be obtained by setting $x_-=p_+=0$ in Eq.~\eqref{Eq:displace} as:
\begin{align}
    \pmb{\hat{b}}_\mr{out} = \tq[b]{T}\,\pmb{\hat{b}}_\mr{in}+\tq[v]{T}^\prime\,\pmb{\hat{v}}_\mr{in}+\frac{x}{x_\mr{SQL}}\,\pmb{t}^\prime, \label{Eq:b_out_disp}
\end{align}
with
\begin{align}
    \tq[b]{T} &= e^{2i\beta}\begin{pmatrix}
        1 & 0 \\
        -\mathcal{K}_\mr{pm} & 1
    \end{pmatrix},\\[1mm]
    \tq[v]{T}^\prime &= e^{2i\beta}\begin{pmatrix}
        0 & 0 \\
        \mathcal{K}_\mr{pm} & 0
    \end{pmatrix}, \\[1mm]
    \pmb{t}^\prime &= e^{i\beta}\begin{pmatrix}
        0 \\
        \sqrt{2\mathcal{K}_\mr{pm}}
    \end{pmatrix}.
\end{align}
Here, \(\mathcal{K}_\mr{pm}\) is the optomechanical coupling constant for the position meter:
\begin{align}
    \mathcal{K}_\mr{pm} = \frac{4\gamma\hbar\,\omega_\mr{a}^2}{m\Omega^2(\gamma^2+\Omega^2)}, \label{Eq.K_pm}
\end{align}
which satisfies the relation:
\begin{align}
    |\mathcal{K}_z|^2 = \mathcal{K}_\mr{sm}\,\mathcal{K}_\mr{pm}.
\end{align}
For simplicity, in the limit \(r\rightarrow \infty\), the measured output \(b_{2,\mr{out}}\) (with \(\phi_\mr{LO}=0\)) is conditioned by combining it with \(\pmb{\alpha}\) using Wiener filters \(g_1\) and \(g_2\):
\begin{align}
    b_{2,\mr{out}}^\mr{con} = b_{2,\mr{out}} + g_1\, x_- + g_2\, p_+. \label{Eq:b_2_out_disp}
\end{align}
The optimal filters are found to be
\begin{align}
    g_1 = e^{2i\beta}\quad \text{and}\quad g_2 = -\mathcal{K}_z^*\,e^{2i\beta},
\end{align}
under which the offline displacement sensitivity in Eq.~\eqref{Eq:b_2_out_disp} agrees with the online value in Eq.~\eqref{Eq:Spec}.

Compared to the online approach or ordinary speed meter schemes, the offline approach has the distinct feature that the back-action force acting on the mirror is not actually canceled. In the toy model of Fig.~\ref{fig:diagram}c, the back-action force exerted by Victor's field on the front side of the mirror has no correlation with the force by Bob's field, so ordinary homodyne detection fails to cancel it. Nevertheless, the back-action is canceled in the eventual sensitivity curve, conditionally erased in post-processing. The key is the Bell measurement, which creates tripartite entanglement among the EPR pair and Victor~\cite{PhysRevLett.80.869}, thereby effectively enabling the cancellation of the back-action force in the offline data. 

\subsection*{Sensitivity analysis}\label{sec.4}
In this section, we first analyze the enhancement in sensitivity relative to both a conventional position meter and the standard quantum limit. We then examine the effect of optical losses, which include power losses at the input port, within the arm cavity, and at the output port. Finally, we apply the proposed technique to interferometric gravitational-wave detectors.

\subsubsection*{Enhancement}
A key feature of the speed meter is its optomechanical coupling factor, \(\mathcal{K}_\mr{sm}\), which remains constant at low frequencies $\Omega\ll\gamma$. Consequently, the speed meter significantly reduces radiation pressure noise. Moreover, since \(\mathcal{K}_\mr{sm}\) remains constant at low frequencies, the radiation pressure noise can be canceled by exploiting correlations between the amplitude \(\hat{b}_\mr{1,out}\) and phase \(\hat{b}_\mr{2,out}\) quadratures without requiring frequency-dependent or variational readout (which would otherwise necessitate additional long filter cavities). The optimal homodyne (readout) angle at DC is given by
\[
\phi_{\mathrm{opt}} = \arccot\Bigl[\mathcal{K}_\mr{sm}(\Omega\rightarrow0)\Bigr].
\]
Thus, with a fixed-angle readout, the sensitivity can beat the SQL in the radiation-pressure–dominated region. In contrast, the coupling constant of the conventional position meter, $\mathcal{K}_\mr{pm}$ shown in Eq.~\eqref{Eq.K_pm}, exhibits a frequency dependence proportional to \(\Omega^{-2}\).

\begin{comment}
\begin{figure}[]
    \centering
    \includegraphics[width=1.0\linewidth]{figures/TSM_lossless_comparison.png}
    \caption{Strain sensitivity of the speed meter versus a conventional Michelson interferometer and the SQL. We use the following parameters: arm length \(L=4\) km, mirror mass \(m=200\) kg, total circulating power \(P_\mr{c}=3\) MW, laser wavelength \(\lambda_p=2\,\mu\)m, and cavity bandwidth of \(\gamma=2\pi\times115\) Hz. The solid red curve corresponds $\phi_\mr{LO}=\phi_\mr{LO,opt}$ and the dashed red curve to $\phi_\mr{LO}=\pi/2$. The blue curve is the sensitivity of the conventional Michelson interferometer and the gray is the SQL. }
    \label{fig:TSM_lossless_comparison}
\end{figure}
\end{comment}

%The second term in Eq.~\eqref{Eq:b_out} arises due to imperfections in the teleportation process and vanishes in the limit of infinite squeezing (\(r\rightarrow\infty\)), where teleportation fidelity approaches unity. The left panel of Fig.~\ref{fig:TSM_combined_comparison} displays the strain sensitivity in this ideal limit. Here, the input Victor state is assumed to be a thermal vacuum (with unit covariance), and the system parameters are chosen to match those in Ref.~\cite{2018LSA.....7...11D}. With \(\phi=\pi/2\), the low-frequency noise nearly traces the SQL, whereas the optimal homodyne angle $\phi=\phi_{\mathrm{opt}}$ allows the sensitivity to beat the SQL---albeit with degraded high-frequency sensitivity due to shot noise. 

\begin{figure}
    \centering
    \includegraphics[width=1\linewidth]{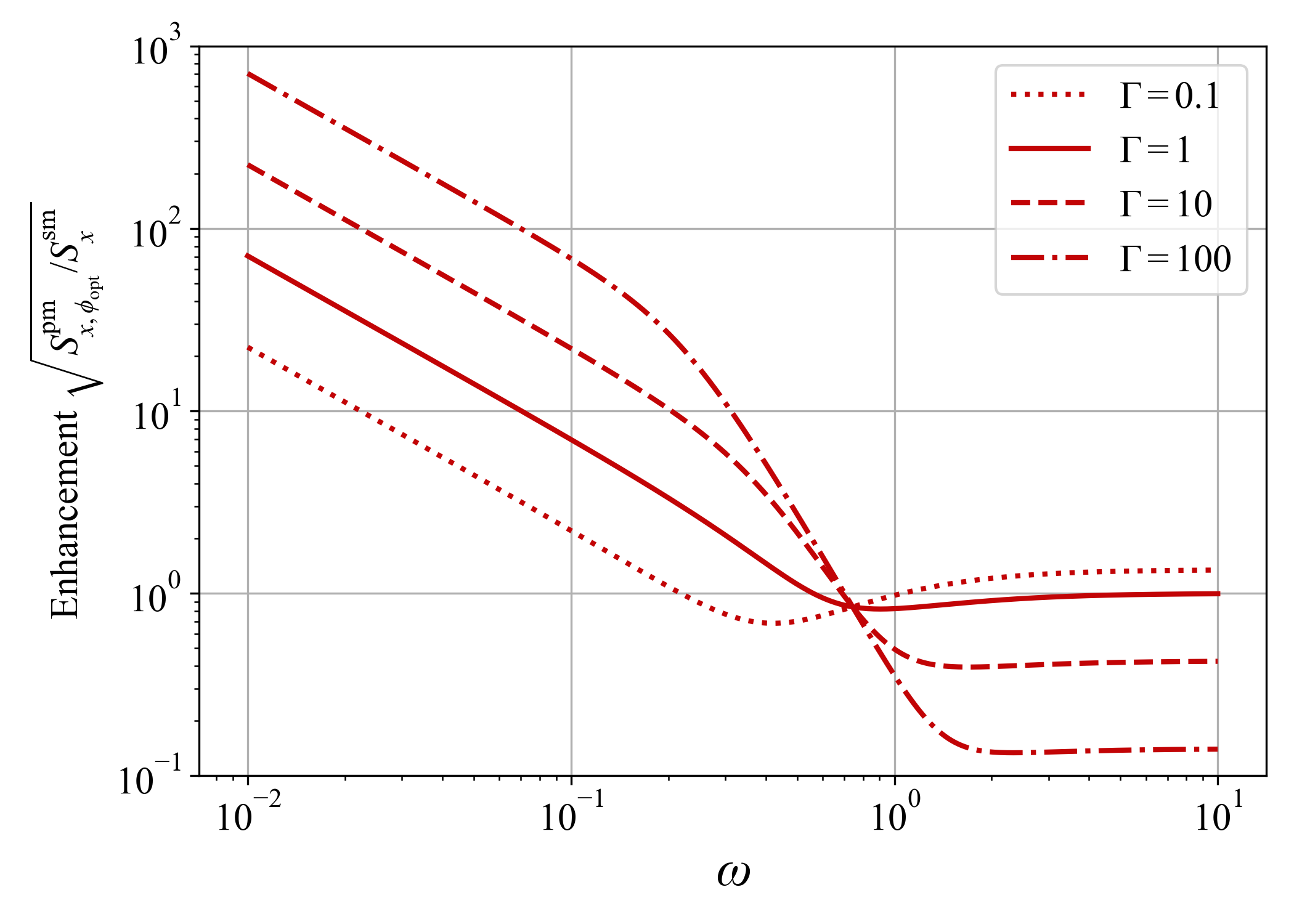}
    \caption{Enhancement in the sensitivity of the speed meter compared to the position meter.}
    \label{fig:enhancement}
\end{figure}

The enhancement factor, in the sensitivity of the speed meter compared to the position meter is given by
\begin{align}
    \frac{S_{x}^\mr{pm}}{S_{x,\phi_\mr{opt}}^\mr{sm}}= \frac{2\omega^2\left[1+\omega^2\left(2+\omega^2\right)\left\{2+(1+\Gamma)\omega^2\left(2+\omega^2\right)\right\}\right]}{(1+\omega^2)\left\{\Gamma+4\omega^4(1+\omega^2)^2\right\}}, \label{eq:enhancement}
\end{align}
where
\begin{align}
    \Gamma = \frac{256 \hbar^2 \omega_a^4}{m^2\gamma^6}, \quad \text{and} \quad
    \omega = \frac{\Omega}{\gamma}.
\end{align}
(for the derivation of the position meter noise $S_{x}^\mr{pm}$, see the supplementary information I).
Equation~\eqref{eq:enhancement} converges to \((1+\Gamma)/2\) in the limit \(\omega \rightarrow \infty\), indicating that the high-frequency (shot-noise–limited) sensitivities of the speed meter and the position meter coincide when \(\Gamma = 1\).
Fig.~\ref{fig:enhancement} shows the enhancement factor for several values of \(\Gamma\). The enhancement at low frequencies becomes more pronounced as \(\Gamma\) increases, which can be realized by increasing the laser power or reducing the mirror mass.

The enhancement beyond the SQL is quantified as
\begin{align}
    \rho^2 = \frac{x_\mr{SQL}^2}{S_{x,\phi_\mr{opt}}^\mr{sm}} = 2\sqrt{\Gamma},
\end{align}
which is equivalent to the parametrization presented in Ref.~\cite{2018LSA.....7...11D}. Thus, the speed meter beats the SQL when \(\Gamma > \frac{1}{4}\).

%We plot the sensitivity of the teleportation speed meter when applied to a LIGO-type gravitational-wave detector, and compare its displacement sensitivity to that of a conventional Michelson interferometer operating with the same total circulating laser power (see the supplementary information I for the derivation of the Michelson interferometer noise). In the radiation-pressure–dominated regime, the speed meter demonstrates superior performance; for example, at 8 Hz its amplitude sensitivity is improved by approximately one order of magnitude.

\subsubsection*{Loss analysis}\label{ssec:lossy case}

We now introduce losses to the sensitivity analysis. For the analysis, we focus on the online approach, but we confirmed that the results are equivalent for the offline case.
Losses inside the arm cavity introduce extra vacuum noise into the equations of motion:
\begin{align}
    -i\Omega \hat{A}_1 &= -(\gamma+\gamma_2)\, \hat{A}_1 + \sqrt{2\gamma}\,\hat{v}_{1,\mr{in}}+\sqrt{2\gamma_2}\hat{a}^\prime_1, \notag\\[1mm]
    -i\Omega \hat{A}_2 &= -(\gamma+\gamma_2)\, \hat{A}_2 + \sqrt{2\gamma}\,\hat{v}_{2,\mr{in}}
    -\sqrt{2}\,\omega_\mr{a}\Bigl(\hat{x}-x\Bigr)\notag\\
    &\qquad +\sqrt{2\gamma_2}\hat{a}^\prime_2, \notag\\[1mm]
    -i\Omega \hat{B}_1 &= -(\gamma+\gamma_2)\, \hat{B}_1 + \sqrt{2\gamma}\,\hat{b}_{1,\mr{in}}+\sqrt{2\gamma_2}\hat{b}^\prime_1, \notag\\[1mm]
    -i\Omega \hat{B}_2 &= -(\gamma+\gamma_2)\, \hat{B}_2 + \sqrt{2\gamma}\,\hat{b}_{2,\mr{in}}
    -\sqrt{2}\,\omega_\mr{a}\Bigl(\hat{x}-x\Bigr) \notag\\
    &\qquad +\sqrt{2\gamma_2}\hat{b}^\prime_1. \label{Eq:EoM3}
\end{align}
Here, the loss-induced vacuum fields ($\pmb{\hat{a}}^\prime=\{\hat{a}^\prime_1, \hat{a}^\prime_2\}^T$ and $\pmb{\hat{b}}^\prime=\{\hat{b}^\prime_1, \hat{b}^\prime_2\}^T$) have unit covariance, and $\gamma_2$ is the mode extraction rate due to the power loss.

Additional losses occur in the detection and input optics. The output fields become:
\begin{align}
    \pmb{\hat{\alpha}}^\mr{L} &= \sqrt{1-\epsilon_\mr{out}}\, \pmb{\hat{\alpha}} + \sqrt{\epsilon_\mr{out}}\, \pmb{\hat{\alpha}}^\prime, \\
    \pmb{\hat{b}}_\mr{out}^\mr{L} &= \sqrt{1-\epsilon_\mr{out}}\, \pmb{\hat{b}}_\mr{out} + \sqrt{\epsilon_\mr{out}}\, \pmb{\hat{b}}^\prime_\mr{out}
\end{align}
and the input Bob's field is modified as:
\begin{align}
    \pmb{\hat{b}}_\mr{in}^\mr{L} &= \sqrt{1-\epsilon_\mr{in}}\, \pmb{\hat{b}}_\mr{in} + \sqrt{\epsilon_\mr{in}}\, \pmb{\hat{b}}^\prime_\mr{in}.
\end{align}
Here $\pmb{\hat{\alpha}}^\prime=\{\hat{\alpha}^\prime_1, \hat{\alpha}^\prime_2\}^T$, $\pmb{\hat{b}}^\prime_\mr{in}=\{\hat{b}^\prime_\mr{1,in}, \hat{b}^\prime_\mr{2,in}\}^T$ and $\pmb{\hat{b}}^\prime_\mr{out}=\{\hat{b}^\prime_\mr{1,out}, \hat{b}^\prime_\mr{2,out}\}^T$ are the loss-induced vacuum fields with unit covariance. Using the transfer matrices for each field, one can compute the overall displacement sensitivity (see the supplementary information II).

%The right panel of Fig.~\ref{fig:TSM_combined_comparison} shows the displacement sensitivity curves in the lossy case. Below 10 Hz, the sensitivity degrades due to uncorrelated vacuum noise introduced by losses. In our simulation, we assumed 1\% input and output losses for each of the three fields and 30 ppm round-trip loss in the arm cavity. The EPR entanglement is generated with 15 dB of squeezing—beyond which the sensitivity no longer improves because the interferometer losses dominate (note that 15 dB is the generated squeezing level, not the observed level). The low-frequency sensitivity scales as $\Omega^{-1}$, reflecting the frequency dependence of the radiation pressure by the loss-induced vacuum. Despite the presence of losses, the low-frequency sensitivity remains enhanced, and with the optimal homodyne angle $\phi_\mr{LO}=\phi_\mr{LO,opt}$ the sensitivity surpasses the SQL.

\subsubsection*{Interferometric gravitational-wave detector}

\begin{figure*}[ht!]
    \centering
    \includegraphics[width=0.7\linewidth]{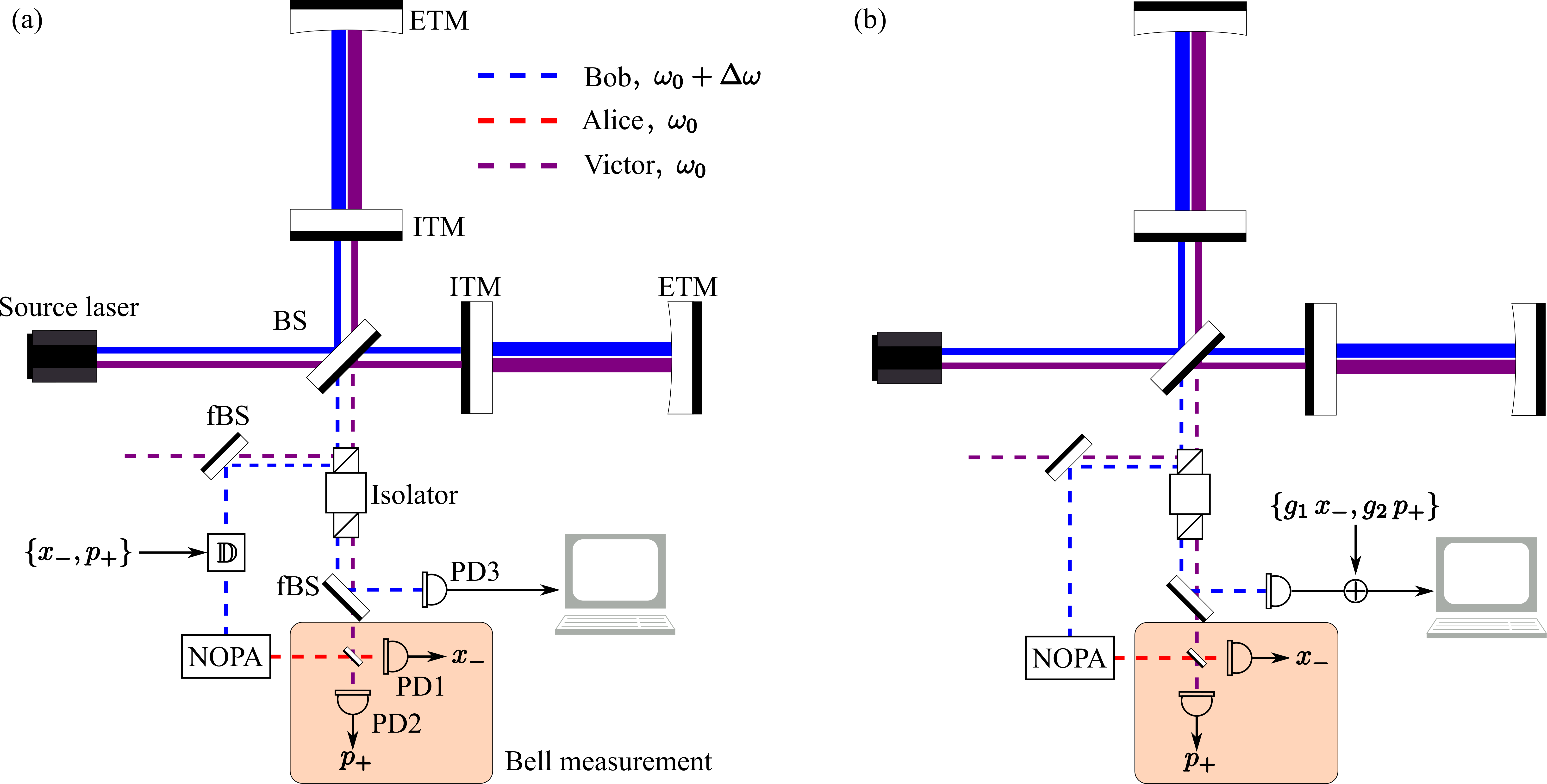}
    \caption{Schematics of the teleportation-based speed meter in a Michelson-type interferometer. 
    (a) Online approach: Victor's field (purple) is combined with Alice's field (red) and sent to a Bell measurement. Based on the measurement outcome, Bob's field (blue) is displaced and then injected into the interferometer; its output is subsequently detected by a homodyne detector. (b) Offline approach: the displacement operation is implemented offline through post-processing. Abbreviations: ITM, input test mass; ETM, end test mass; BS, beam splitter; fBS, frequency beam splitter; NOPA, non-degenerate optical parametric amplifier; PD, photodetector.}
    \label{fig:TeleportationSpeedmeter}
\end{figure*}

We apply the teleportation-based speed measurement scheme to a LIGO-type gravitational-wave detector and compare its displacement sensitivity to that of a conventional Michelson interferometer operating as a position meter, both using the same total circulating laser power. Panels (a) and (b) of Fig.~\ref{fig:TeleportationSpeedmeter} show the schematics of the online and offline approaches in a Michelson-type interferometer, respectively. The input fields, Bob and Victor, are injected through the Faraday isolator. Alice’s and Victor’s fields share the same frequency, \(\omega_0\), while Bob's frequency is slightly detuned to \(\omega_0+\Delta\omega\). The pump fields at frequencies \(\omega_0\) and \(\omega_0+\Delta\omega\) co-resonate in the arm cavities. The beam splitters at the dark port combining the three beams are frequency-dependent; for example, a triangular optical cavity can be used to realize this. The coupling constant \(\omega_\mr{a}\) is defined as
\[
\omega_\mr{a} \equiv \sqrt{\frac{mJ}{2\hbar}},
\]
with the normalized power \(J=\frac{4 P_c \omega_0}{m L c}\), where \(P_c\) is the total circulating power, \(\omega_0\) is the laser frequency, and \(c\) is the speed of light. In the online approach, a displacement operation is applied at the dark port, whereas in the offline approach all operations are implemented via post-processing.

\begin{figure*}[ht!]
    \centering
    \includegraphics[width=0.8\linewidth]{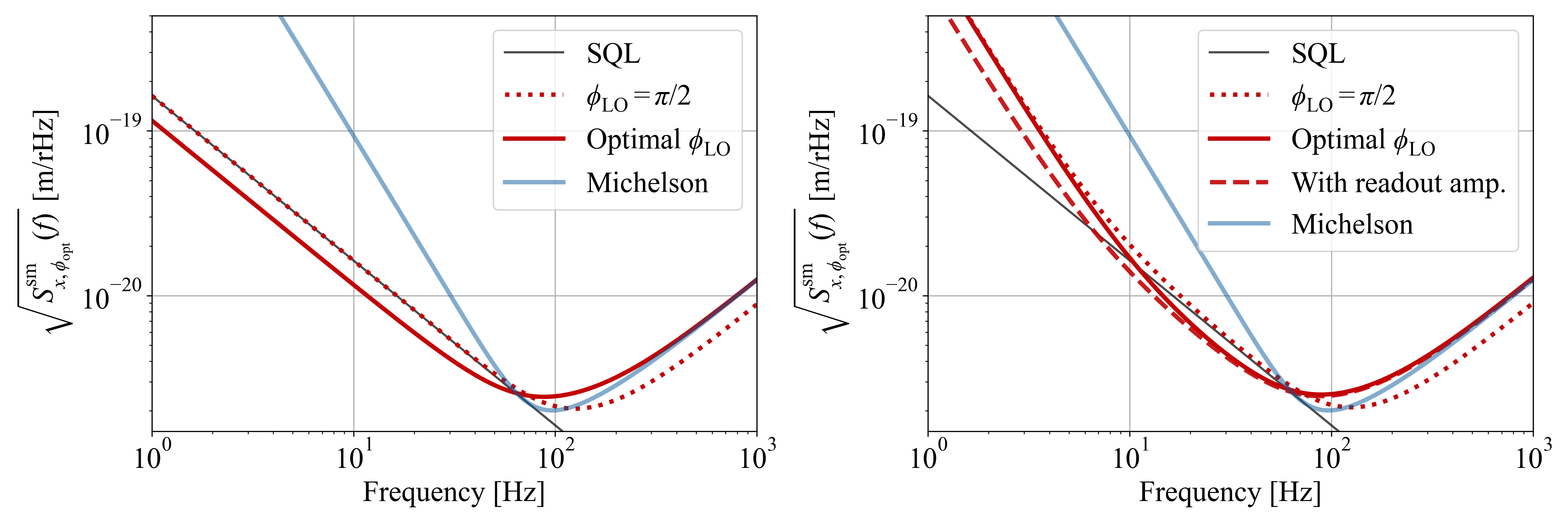}
    \caption{Displacement sensitivity of the speed meter applied to a gravitational--wave detector. Left panel: Sensitivity in the lossless limit and the SQL. Right panel: Sensitivity with losses. The dashed curve in the right panel is the sensitivity with a readout amplifier. The blue curve in both panels is the sensitivity of a conventional Michelson interferometer without loss or input squeezing. In the right panel, we assume input/output losses of $\epsilon_\mr{in}=\epsilon_\mr{out}=1\%$, an arm cavity loss of 30 ppm, and 15 dB of generated two-mode squeezing of Alice and Bob (Victor is vacuum). For both panels, we use the following parameters: arm length \(L=4\) km, mirror mass \(m=200\) kg, total circulating power \(P_\mr{c}=3\) MW, laser wavelength \(\lambda_p=2\,\mu\)m, and cavity bandwidth of \(\gamma=2\pi\times115\) Hz~\cite{2018LSA.....7...11D}. The solid red curve corresponds $\phi_\mr{LO}=\phi_\mr{LO,opt}$ and the dotted red curve to $\phi_\mr{LO}=\pi/2$.}
    \label{fig:TSM_combined_comparison}
\end{figure*}

The left panel of Fig.~\ref{fig:TSM_combined_comparison} shows the displacement sensitivity curves in the lossy case. The system parameters are chosen to match those in Ref.~\cite{2018LSA.....7...11D}, where the shot-noise–limited sensitivity of the position meter and the speed meter with the optimal homodyne angle are aligned to the same level by setting \(\Gamma \sim 1\).
At \(\phi=\pi/2\), the low-frequency noise nearly reaches the SQL, whereas using the optimal homodyne angle \(\phi=\phi_{\mathrm{opt}}\) allows the sensitivity to surpass the SQL. Compared to the position meter, the amplitude sensitivity is improved by approximately one order of magnitude at 8 Hz.

The right panel of Fig.~\ref{fig:TSM_combined_comparison} shows the displacement sensitivity curves in the lossy case. Although losses degrade sensitivity, our proposed scheme still yields a considerable enhancement at low frequencies.
Below 10 Hz, the sensitivity degrades due to uncorrelated vacuum noise introduced by losses. In our simulation, we assumed 1\% input and output losses for each of the three fields and $\mathcal{L}=30$ ppm round-trip loss in the arm cavity. The arm loss contributes as the additional damping $\gamma_2$ as $\gamma_2=\frac{\mathcal{L}c}{4L}$. The EPR entanglement is generated with 15 dB of squeezing—beyond which the sensitivity no longer improves because the interferometer losses dominate (note that 15 dB is the generated squeezing level, not the observed level). The low-frequency sensitivity scales as $\Omega^{-1}$, reflecting the frequency dependence of the radiation pressure by the loss-induced vacuum. Despite the presence of losses, the low-frequency sensitivity remains enhanced, and with the optimal homodyne angle $\phi_\mr{LO}=\phi_\mr{opt}$ the sensitivity surpasses the SQL.

\begin{comment}
\begin{figure}[]
    \centering
    \includegraphics[width=1.0\linewidth]{figures/TSM_lossy_comparison.png}
    \caption{Strain sensitivity with losses. Colors and line styles follow Fig.~\ref{fig:TSM_combined_comparison}. We assume input/output losses of $\epsilon_\mr{in}=\epsilon_\mr{out}=1\%$, an arm cavity loss of 30 ppm, and 15 dB of generated two-mode squeezing.}
    \label{fig:TSM_lossy_comparison}
\end{figure}
\end{comment}

\section*{Discussion}\label{sec.5}
In the online approach, the teleportation works as a frequency converter, which is analogous to the polarization circulator in Ref.~\cite{2018LSA.....7...11D}. The benefit of using quantum teleportation is that this operation can instead be achieved offline, which is not possible with a classical frequency converter.

When applied to interferometric displacement sensors, our scheme requires no modifications to the core interferometer optics including the coating properties only by altering the dark-port components and introduces a second pump from the bright port (which is similar to the paired carrier interferometry in Ref.~\cite{PhysRevD.78.062003, PhysRevD.91.042004}). The teleportation-based speed meter needs only two pumping lasers at slightly different frequencies (detuned on the order of MHz). Such two-color lasers can be generated, for example, by using an electro-optic modulator.
%Unlike polarization-based speed meters~\cite{2018LSA.....7...11D}—which require identical optical properties for both $s$ and $p$ polarizations—the teleportation-based speed meter needs only two pumping lasers at slightly different frequencies (detuned on the order of MHz). Such two-color lasers can be generated, for example, by using an electro-optic modulator.

%Furthermore, the teleportation-based speed meter is compatible with other entanglement-based schemes~\cite{Ma_2017,PhysRevA.110.022601}. Squeezing the input Victor state can further improve sensitivity. However, with a fixed squeezing angle, the shot noise worsens near and above $\Omega\sim\gamma$ because $\mathcal{K}_\mr{sm}$ is no longer constant. One potential approach, without requiring additional filter cavities, is to generate EPR-type entanglement between Victor and a third party (called Charles) and inject these fields into the differential port of the interferometer. (Here, while Victor and Bob are pumped from the common port, Charles is not.) If the detector configuration allows tuning the bandwidth and detuning for Charles (e.g., a signal-recycled Fabry-Pe\'rot scheme), one can realize an EPR-broadband speed meter.

The simulation presented omits some experimental details of the teleportation procedure. In the online approach, the physical displacement operation, using a low-transmissivity mirror and coherent laser with modulation~\cite{doi:10.1126/science.282.5389.706}, may introduce additional losses, making the offline method more favorable in practice.

A major challenge in entanglement-based schemes is to mitigate the noise contribution from input and output losses. In our approach, three detection ports contribute output noise (a threefold effect), while input losses affect both Victor and Bob (a twofold effect). The readout noise can be mitigated by employing a readout amplifier~\cite{PhysRevD.26.1817,Knyazev:19, Frascella_2021,Kwan_2024}. The dashed curve in the right panel of Fig.~\ref{fig:TSM_combined_comparison} shows that readout amplification improves sensitivity in the radiation-pressure–dominated band, compared to that without the amplifier.% (which is identical to the solid red curve in Fig.~\ref{fig:TSM_combined_comparison}). 
Readout amplification is crucial not only for the teleportation-based speed meter but also for other multi-color interferometric techniques.

%We conclude this section with a brief discussion of the scientific impact. Since our approach is based on the polarization-type speed meter~\cite{2018LSA.....7...11D}, one promising scientific impact is the observation rate of intermediate mass black hole mergers. As shown in Ref.~\cite{2018LSA.....7...11D}, the event rate for the mergers between black holes with $10^2$--$10^3$ solar masses improves by more than a factor of 100 using a speed meter compared to a conventional Michelson-type detector.

\begin{comment}
\begin{figure}[h]
    \centering
    \includegraphics[width=1.0\linewidth]{figures/TSM_no_out_loss_comparison.png}
    \caption{Strain sensitivity with and without readout amplification. The dashed-dot is the sensitivity with a readout amplifier. The solid red is the same curve as that in Fig.~\ref{fig:TSM_lossy_comparison}.}
    \label{fig:TSM_no_out_loss_comparison}
\end{figure}
\end{comment}

%\section{Conclusion}\label{sec.6}
In this article, we propose an alternative speed meter scheme for interferometric displacement sensors based on quantum teleportation. We introduce two equivalent implementations—an online approach and an offline approach—both of which achieve quantum noise reduction below the standard quantum limit (SQL), even in the presence of losses. When applied to interferometric gravitational-wave detectors, the scheme requires no modification to the core interferometer design or to mirror properties such as coatings or substrate materials. This provides a practical route to realizing speed meter operation and improving low-frequency sensitivity.

\begin{acknowledgments}
Research by Y. N. is supported by JSPS Grant-in-Aid for JSPS Fellows Grant Number 23KJ0787 and 23K25901. J.W.G. is supported by the Australian Research Council Centre of Excellence for Gravitational Wave Discovery (Project No. CE170100004 and CE230100016), an Australian Government Research Training Program Scholarship, and also partially by the US NSF grant PHY-2011968. In addition, Y.C. acknowledges the support by the Simons Foundation (Award Number 568762). This work was partially supported by JST ASPIRE (JPMJAP2320). 
\end{acknowledgments}

%\input{chap_}
%\input{chap_notations_values}
%\includepdf[pages=-]{hand_calculation.pdf}

\bibliography{ref} %
%-------------------

\end{document}